\documentclass{jfm}
\usepackage{epsfig}

\ifCUPmtlplainloaded \else
  \checkfont{eurm10}
  \iffontfound
    \IfFileExists{upmath.sty}
      {\typeout{^^JFound AMS Euler Roman fonts on the system,
                   using the 'upmath' package.^^J}%
       \usepackage{upmath}}
      {\typeout{^^JFound AMS Euler Roman fonts on the system, but you
                   dont seem to have the}%
       \typeout{'upmath' package installed. JFM.cls can take advantage
                 of these fonts,^^Jif you use 'upmath' package.^^J}%
      }
  \else
  \fi
\fi

\ifCUPmtlplainloaded \else
  \checkfont{msam10}
  \iffontfound
    \IfFileExists{amssymb.sty}
      {\typeout{^^JFound AMS Symbol fonts on the system, using the
                'amssymb' package.^^J}%
       \usepackage{amssymb}%
         \let\leq=\leqslant
         \let\geq=\geqslant
      }{}
  \fi
\fi

\ifCUPmtlplainloaded \else
  \IfFileExists{amsbsy.sty}
    {\typeout{^^JFound the 'amsbsy' package on the system, using it.^^J}%
     \usepackage{amsbsy}}
    {\providecommand\boldsymbol[1]{\mbox{\boldmath $##1$}}}
\fi

\newcommand\xib{\boldsymbol{\xi}}
\newcommand\psib{\boldsymbol{\psi}}
\newcommand\Gb{\boldsymbol{\Gamma}}
\newcommand\bchi{\boldsymbol{\chi}}

\newsavebox{\astrutbox}
\sbox{\astrutbox}{\rule[-5pt]{0pt}{20pt}}

\newcommand\p{\ensuremath{\partial}}

\newcommand{\dint}{\displaystyle\int}

\newcommand{\dfrac}{\displaystyle\frac}

\def\a{\alpha}
\def\b{{\bf b}}
\def\c{{\bf c}}
\def\d{\delta}
\def\e{\varepsilon}
\def\eps{\epsilon}

\def\q{{\bf q}}
\def\k{{\bf k}}
\def\x{{\bf x}}
\def\y{{\bf y}}
\def\l{\ell}

\def\p{{\bf p}}
\def\s{\sigma}
\def\v{\varphi}
\def\D{\Delta}

\def\L{{\bf L}}

\def\U{{\bf U}}
\def\V{{\bf V}}
\def\W{{\bf W}}
  
\newcommand{\gbout}[1]{} 

\begin{document}

\title[Water wave transport]{Capillary-gravity wave transport 
over spatially random drift}

\author[Bal \& Chou]
{G\ls U\ls I\ls L\ls L\ls A\ls U\ls M\ls E \ns B\ls A\ls L$^{*}$ \ns 
and \ns T\ls O\ls M C\ls H\ls O\ls U$^{\dagger}$}

\affiliation{$^{*}$ Department of Mathematics,
  University of Chicago,
  Chicago, IL 60637 \\
  $^{\dagger}$Department of Mathematics, Stanford University,
  Stanford, CA 94305}

\maketitle

\begin{abstract}

We derive transport equations for the propagation of water wave action in the
presence of a static, spatially random surface drift.  Using the Wigner
distribution $\W(\x,\k,t)$ to represent the envelope of the wave amplitude at
position $\x$ contained in waves with wavevector $\k$, we describe surface
wave transport over static flows consisting of two length scales; one varying
smoothly on the wavelength scale, the other varying on a scale comparable to
the wavelength.  The spatially rapidly varying but weak surface flows
augment the characteristic equations with scattering terms that are explicit
functions of the correlations of the random surface currents. These
scattering terms depend parametrically on the magnitudes and directions of
the smoothly varying drift and are shown to give rise to a Doppler coupled
scattering mechanism.  The Doppler interaction in the presence of slowly
varying drift modifies the scattering processes and provides a mechanism for
coupling long wavelengths with short wavelengths.  Conservation of wave
action (CWA), typically derived for slowly varying drift, is extended to
systems with rapidly varying flow.  At yet larger propagation distances, we
derive from the transport equations, an equation for wave energy diffusion.
The associated diffusion constant is also expressed in terms of the surface
flow correlations.  Our results provide a formal set of equations to analyse
transport of surface wave action, intensity, energy, and wave scattering as a
function of the slowly varying drifts and the correlation functions of the
random, highly oscillatory surface flows.


\end{abstract}

\section{Introduction}


Water wave dynamics are altered by interactions with spatially varying
surface flows.  The surface flows modify the free surface boundary conditions
that determine the dispersion for propagating water waves.  The effect of
smoothly varying (compared to the wavelength) currents have been analysed
using ray theory (\cite{PEREGRINE76,JONSSON90}) and the principle of
conservation of wave action (CWA) (cf.  \cite{LH,MEI,WHITE,WHITHAM} and
references within).  These studies have largely focussed on long ocean
gravity waves propagating over even larger scale spatially varying drifts. 
Water waves can also scatter from regions of underlying vorticity regions
smaller than the wavelength \cite{FABRIKANT} and \cite{LUND}.  Boundary
conditions that vary on capillary length scales, as well as wave interactions
with structures comparable to or smaller than the wavelength can also lead to
wave scattering (\cite{CHOUB,GOU}), attenuation (\cite{CHOUA,KYL}), and Bragg
reflections (\cite{CHOUC,MEI1}).  Nonetheless, water wave propagation over
random static underlying currents that vary on both large and small length
scales, and their interactions, have received relatively less attention.



In this paper, we will only consider static irrotational currents, but
derive the transport equations for surface waves in the presence of
underlying flows that vary on {\it both} long and short (on the order
of the wavelength) length scales.  Rather than computing wave
scattering from specific static flow configurations
(\cite{GERBER,TRULSON,FABRIKANT}), we take a statistical approach by
considering ensemble averages over realisations of the static
randomness.  Different statistical approaches have been applied to
wave propagation over a random depth (\cite{ELTER}), third sound
localization in superfluid Helium films (\cite{SUPERF}), and wave
diffusion in the presence of turbulent flows
(\cite{HOWE,RUSSKI,RYZHIK}).

In the next section we derive the linearised capillary-gravity wave
equations to lowest order in the irrotational surface flow.  The fluid
mechanical boundary conditions are reduced to two partial differential
equations that couple the surface height to velocity potential at the
free surface.  We treat only the ``high frequency'' limit (\cite{RPK})
where wavelengths are much smaller than wave propagation distances
under consideration. In Section 3, we introduce the Wigner
distribution ${\bf W}(\x,\k,t)$ which represents the wave energy
density and allows us to treat surface currents that vary
simultaneously on two separated length scales.  The dynamical
equations developed in section 2 are then written in terms of an
evolution equation for ${\bf W}$.  Upon expanding ${\bf W}$ in powers
of wavelength/propagation distance, we obtain transport equations.

In Section 4, we present our main mathematical result, equation
(\ref{RESULT1}), an equation describing the transport of surface wave
action.  Appendix A gives details of some of the derivation. The
transport equation includes advection by the slowly varying drift,
plus scattering terms that are functions of the correlations of the
rapidly varying drift, representing water wave scattering. Upon
simultaneously treating both smoothly varying and rapidly varying
flows using a two-scale expansion, we find that scattering from
rapidly varying flows depends parametrically on the smoothly varying
flows.  In the Results and Discussion, we discuss the regimes of
validity, consider specific forms for the correlation functions, and
detail the conditions for doppler coupling.  CWA is extended to
include rapidly varying drift provided that the correlations of the
drift satisfy certain constraints. We also physically motivate the
reason for considering two scales for the underlying drift. In the
limit of still larger propagation distances, after multiple wave
scattering, wave propagation leaves the transport regime and becomes
diffusive (\cite{SHENG}).  A diffusion equation for water wave energy
is also given, with an outline of its derivation given in Appendix B.

\section{Surface wave equations}\label{sec:eqns}

Assume an underlying flow $\V(\x,z)\equiv(U_{1}(\x,z), U_{2}(\x,z),
U_{z}(\x,z)) \equiv (\U(\x,z), U_{z}(\x,z))$, where the 1,2 components
denote the two-dimensional in-plane directions.  This static flow may
be generated by external, time independent sources such as wind or
internal flows beneath the water surface. The surface deformation due
to $\V(\x,z)$ is denoted $\bar{\eta}(\x)$ where $\x \equiv (x,y)$ is
the two-dimensional in-plane position vector.  An additional variation
in height due to the velocity ${\bf v}(\x,z)$ associated with surface
waves is denoted $\eta(\x,t)$.  When all flows are irrotational, we
can define their associated velocity potentials
$\V(\x,z)\equiv(\nabla_{\x}+\hat{{\bf z}}\partial_{z})\Phi(\x,z)$ and
${\bf v}(\x,z,t)\equiv(\nabla_{\x}+\hat{{\bf
    z}}\partial_{z})\v(\x,z,t)$.  Incompressibility requires
\begin{equation}
\D\v(\x,z,t)+
\partial_{z}^{2}\v(\x,z,t) = \D\Phi(\x,z)+
\partial_{z}^{2}\Phi(\x,z)=0,
\label{LAPLACE}
\end{equation}
where $\D=\nabla_{\x}^{2}$ is the two-dimensional Laplacian. The
kinematic condition (\cite{WHITHAM}) applied at
$z=\bar{\eta}(\x)+\eta(\x,t)\equiv \zeta(\x,t)$ is
\begin{equation}
\partial_{t}\eta(\x,t)+\U(\x, \zeta)\cdot\nabla_{\x}\zeta(\x,t)
= U_{z}(\x,z=\zeta)+\partial_{z}\v(\x,z=\zeta,t).
\label{KINEMATIC}
\end{equation}
Upon expanding (\ref{KINEMATIC}) to linear order in $\eta$ and $\v$
about the static free surface, the right hand side becomes
\begin{equation}
U_{z}(\x,\zeta)+\partial_{z}\v(\x,\zeta,t) 
= U_{z}(\x,\bar{\eta})+\eta(\x,t)\partial_{z}U_{z}(\x,\bar{\eta}) 
+ \partial_{z}\v(\x,\bar{\eta},t) + O(\eta^{2}).
\label{KINEMATIC2}
\end{equation}
At the static surface $\bar{\eta}$, $\U(\x,
\bar{\eta})\cdot\nabla_{\x}\bar{\eta}(\x) = U_{z}(\x,\bar{\eta})$.
Now assume that the underlying flow is weak enough such that
$U_{z}(\x,z\approx 0)$ and $\bar{\eta}$ are both small. A rigid
surface approximation is appropriate for small Froude numbers
$U^{2}/c^{2}_{\phi}\sim \vert\nabla_{\x}\bar{\eta}\vert^{2} \sim
U_{z}(\x,0)/\vert \U(\x,0)\vert \ll 1$ ($c_{\phi}$ is the surface wave
phase velocity) when the free surface boundary conditions can be
approximately evaluated at $z=0$ (\cite{FABRIKANT}). Although we have
assumed $U_{z}(\x,z\approx 0)=\partial_{z}\Phi(\x,z\approx 0) \approx
0$ and a vanishing static surface deformation $\bar{\eta}(\x)\approx
0$, $\nabla_{\x}\cdot\U(\x,0) = -\partial_{z}U_{z}(\x,0) \neq 0$.

Combining the above approximations with the dynamic boundary
conditions (derived from balance of normal surface stresses at $z=0$
(\cite{WHITHAM})), we have the pair of coupled equations
\begin{equation}
\begin{array}{l}
\displaystyle 
\partial_{t}\eta(\x,t)+\nabla_{\x}\cdot\left(\U(\x,z=0)\eta(\x,t)\right) 
= \lim_{z\rightarrow 0^{-}}
\partial_{z}\v(\x,z,t)\\[13pt]
\displaystyle 
\lim_{z\rightarrow 0^{-}}\left[\rho\partial_{t}\v(\x,z,t)
+\rho{\bf U}(\x,z)\cdot
\nabla_{\x}\v(\x,z,t)\right] =  
\sigma\D \eta(\x,t)-\rho
g\eta(\x,t)
\end{array}
\label{BC}
\end{equation}
where $\sigma$ and $g$ are the air-water surface tension and
gravitational acceleration, respectively.  Although it is
straightforward to expand to higher orders in $\bar{\eta}(\x)$ and
$\eta(\x,t)$, or to include underlying vorticity, we will limit our
study to equations (\ref{BC}) in order to make the development of the
transport equations more transparent.
\begin{center}
\begin{figure}
\epsfig{file=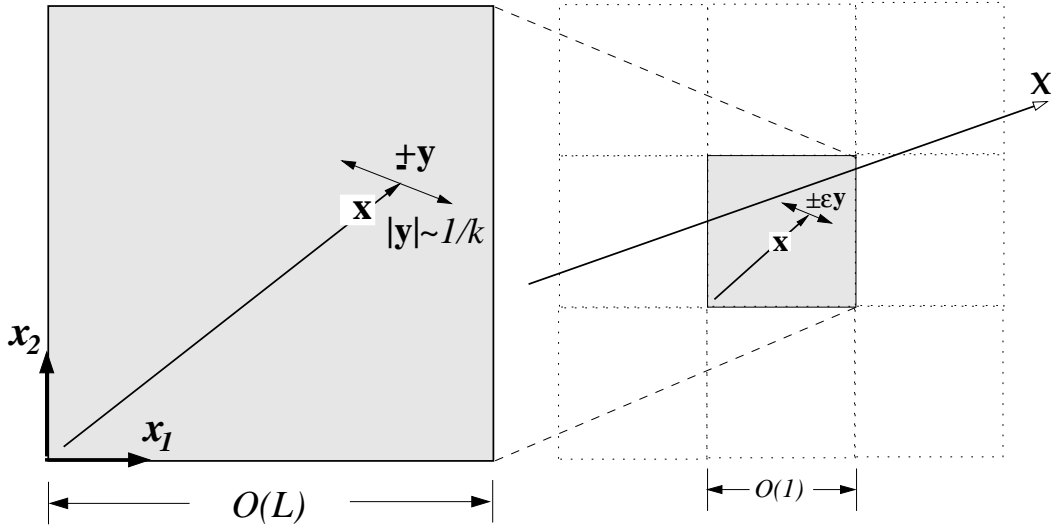,height = 7cm}
\caption{The relevant scales in water wave transport. Initially, the system
  size, observation point, and length scale of the slowly varying
  drift is $O(L)$, with surface wave wavelength and scale of the
  random surface current of $O(1)$.  Upon rescaling, the system size
  becomes $O(1)$, while the wavelength and random flow variations
  are $O(\e)$.}
\label{FIG1}
\end{figure} 
\end{center}
The typical system size, or distance of wave propagation shown in Fig.
\ref{FIG1} is of $O(L)$ with $L\gg 1$. Wavelengths however, are of
$O(1)$.  To implement our high frequency (\cite{RPK}) asymptotic
analyses, we rescale the system such that all distances are measured
in units of $L \equiv \e^{-1}$.  We eventually take the limit $\e
\rightarrow 0$ as an approximation for small, finite $\e$. Surface
velocities, potentials, and height displacements are now functions of
the new variables $\x \rightarrow \x/\e, z\rightarrow z/\e$ and
$t\rightarrow t/\e$.  We shall further nondimensionalise all distances
in terms of the capillary length $\l_{c} = \sqrt{\sigma/g\rho}$.
Time, velocity potentials, and velocities are dimensionalised in units
of $\sqrt{\l_{c}/g}, \sqrt{g\l_{c}^{3}}$, and $\sqrt{g\l_{c}}$
respectively, {\it e.g.} for water, $U=1$ corresponds to a surface
drift velocity of $\sim 16.3$cm/s.

Since $U_{z}(\x,z\approx 0)\approx 0$, we define the flow at the
surface by
\begin{equation}
\U(\x,z=0) \equiv \U(\x)+\sqrt{\e}\d\U(\x/\e). 
\end{equation}
In these rescaled coordinates, $\U(\x)$ denotes surface flows varying
on length scales of $O(1)$ much greater than a typical wavelength,
while $\d\U(\x/\e)$ varies over lengths of $O(\e)$ comparable to a
typical wavelength. The {\it amplitude} of the slowly varying flow
$\U(\x)$ is $O(\e^{0})$, while that of the rapidly varying flow
$\d\U(\x/\e)$, is assumed to be of $O(\sqrt{\e})$.  A more detailed
discussion of the physical motivation for considering the $\sqrt{\e}$
scaling is deferred to the Results and Discussion. After rescaling,
the boundary conditions (\ref{BC}) evaluated at $z=0$ become
\begin{equation}
\begin{array}{l}
\displaystyle 
\partial_{t}\eta(\x,t)+\nabla_{\x}\cdot\left[\left(\U(\x)+
\sqrt{\e}\delta\U(\x/\e)\right)
\eta(\x)\right] = \lim_{z\rightarrow 0^{-}}\partial_{z}\v(\x,0)\\[13pt]
\displaystyle \partial_{t}\v(\x,t)
+ {\bf U}(\x)\cdot
\nabla_{\x}\v(\x,t) + \sqrt{\e}\d\U(\x/\e)\cdot
\nabla_{\x}\v(\x,t)=  \e \D \eta(\x,t)
-\e^{-1}\eta(\x,t).
\end{array}
\label{BC2}
\end{equation}
Although drift that varies slowly along one wavelength can be treated
with characteristics and WKB theory, random flows varying on the
wavelength scale require a statistical approach.  Without loss of
generality, we choose $\d\U$ to have zero mean and an isotropic
two-point correlation function $\langle\d U_{i}(\x)\d
U_{j}(\x')\rangle \equiv R_{ij}(\vert\x-\x'\vert)$, where
$(i,j)=(1,2)$ and $\langle\ldots\rangle$ denotes an ensemble average
over realisations of $\d\U(\x)$.

We now define the spatial Fourier decompositions for the dynamical
wave variables
\begin{equation}
\begin{array}{ll}
\displaystyle \v(\x,-h\leq z\leq \zeta,t) = 
\int_{\q} \, \v(\q,t) 
e^{-i\q\cdot\x}{\cosh q(h+z)\over
\cosh qh}, & \displaystyle 
\eta(\x,t) = \int_{\q} \,\eta(\q,t) e^{-i\q\cdot\x},
\end{array}
\label{FTETAPHI}
\end{equation}
the static surface flows
\begin{equation}
\begin{array}{ll}
\displaystyle {\bf U}(\x) = 
\int_{\q} {\bf U}(\q) e^{-i\q\cdot\x}, & \displaystyle 
\d\U\left({\x\over \e}\right)
=\int_{\q} \d\U(\q) e^{-i\q\cdot\x/\e},
\end{array}
\label{FTU}
\end{equation}
and the correlations
\begin{equation}
R_{ij}(\x) = 
\int_{\q} R_{ij}(\q) e^{-i\q\cdot\x},  
\label{FTR}
\end{equation}
where $\q = (q_{1}, q_{2})$ is an in-plane two dimensional wavevector,
$q \equiv \vert\q\vert=\sqrt{q_{1}^{2}+q_{2}^{2}}$, and
$\int_{\q}\equiv (2\pi)^{-2}\int dq_{1}dq_{2}$.  The Fourier integrals
for $\eta$ exclude $\q=0$ due to the incompressibility constraint
$\int_\x \eta(\x,t) = 0$, while the $\q=0$ mode for $\v$ gives an
irrelevant constant shift to the velocity potential.  Note that $\v$
in (\ref{FTETAPHI}) manifestly satisfies (\ref{LAPLACE}). Substituting
(\ref{FTU}) into the boundary conditions (\ref{BC}), we obtain,
\begin{equation}
\begin{array}{l}
\displaystyle \partial_{t}\eta(\k,t) - 
i\int_{\q}\eta(\k-\q) {\bf U}(\q)\cdot\k
- i\sqrt{\e}\int_{\q}\eta(\k-\q/\e)
\d\U(\q)\cdot\k = \v(\k,t)\,k\tanh \e kh \\[13pt]
\displaystyle \partial_{t}\v(\k,t) - i\int_{\q} 
{\bf U}(\q)\cdot\left(\k-\q\right)\v(\k-\q)
- i\sqrt{\e}\int_{\q} 
\d\U(\q)\cdot\left(\k-\q/\e\right)
\v(\k-\q/\e) \\[13pt]
\hspace{3in} = -(\e k^2+\e^{-1})\eta(\k).
\end{array}
\label{FTBC}
\end{equation}
where the $\d\U(\q)$ are correlated according to
\begin{equation}
\langle\d U_{i}(\p)\d U_{j}(\q)\rangle
=R_{ij}(\vert\p\vert)\d(\p+\q).
\label{CORRELATION}
\end{equation}
Since the correlation $R_{ij}(\x)$ is symmetric in $i\leftrightarrow
j$, and depends only upon the magnitude $\vert\x\vert$,
$R_{ij}(\vert\p-\q\vert)$ is real.

In the case where $\d\U=0$ and $\U(\x)\equiv \U_{0}$ is strictly
uniform, equations (\ref{FTBC}) can be simplified by assuming a
$e^{-i\omega t}$ dependence for all dynamical variables.  Uniform
drift yields the familiar capillary-gravity wave dispersion relation
\begin{equation}
\omega(\k) = \sqrt{(k^{3}+k)\tanh kh}+\U_{0}\cdot\k \equiv
\Omega(\k)+\U_{0}\cdot\k.
\end{equation}
However, for what follows, we wish to derive transport equations for
surface waves (action, energy, intensity) in the presence of a
spatially varying drift containing two length scales:
$\U=\U(\x)+\sqrt{\e}\d\U(\x/\e)$.

\section{The Wigner distribution and asymptotic analyses}

The intensity of the dynamical wave variables can be represented by the
product of two Green functions evaluated at positions $\x\pm \e\y/2$. The
difference in their evaluation points, $\e\y$, resolves the waves of
wavevector $\vert\k\vert \sim 2\pi/(\e y)$. \cite{ELTER} used this
representation to study shallow water wave propagation over a random bottom. 
However, for the arbitrary depth surface wave problem, where the Green
function is not simple, and where two length scales are treated, it is
convenient to use the Fourier representation of the Wigner distribution
(\cite{WIGNER,GMMP,RPK}).

Define $\psib=(\psi_{1},\psi_{2})\equiv (\eta(\x),
\v(\x,z=0))$ and the Wigner distribution:
\begin{equation}
\label{eq:wigner}
W_{ij}(\x, \k,t) \equiv (2\pi)^{-2}\int 
e^{i\k\cdot\y}\psi_{i}\left(\x-{\e\y\over 2},t\right)
\psi^{*}_{j}\left(\x+{\e\y\over 2},t\right) d\y
\end{equation}
where $\x$ is a central field point from which we consider two
neighbouring points $\x\pm{\e\y\over 2}$, and their intervening wave
field.  Fourier transforming the $\x$ variable using the definition
(\ref{FTETAPHI}) we find,
\begin{equation}
W_{ij}(\p,\k,t) = (2\pi\e)^{-2}\psi_{i}\left({\p\over 2}-{\k\over
\e},t\right)\psi^{*}_{j}
\left(-{\p\over 2}-{\k\over \e},t\right).
\label{WIGNERP}
\end{equation}
The total wave energy, comprising gravitational, kinetic, and surface
tension contributions is
\begin{equation}
\begin{array}{rl}
{\cal E} = & \displaystyle {1\over 2}\int_\x
\left[\vert\nabla_{\x}\eta(\x)\vert^{2}+
\vert\eta(\x)\vert^{2}\right] +
{1\over 2}\int_\x\int_{-h}^{0} dz \vert
\U(\x,z)+{\bf \hat{z}}U_{z}(\x,z)
+{\bf v}(\x,z)\vert^{2} \\[13pt]
\: & \displaystyle  \hspace{2cm}
-{1\over 2}\int_\x\int_{-h}^{0} dz\vert
\U(\x,z)+{\bf \hat{z}}U_{z}(\x,z)\vert^{2}\\[13pt]
\: & \displaystyle = {1\over 2}\int_{\k}
(k^{2}+1)\vert\eta(\k)\vert^{2} + k\tanh kh
\vert\v(\k,z=0)\vert^{2}.
\end{array}
\label{ENERGY}
\end{equation}
The energy above has been expanded to an order in $\eta(\x,t)$ and
$\v(\x,z,t)$ consistent with the approximations used to derive
(\ref{BC}).  In arriving at the last equality in (\ref{ENERGY}), we
have integrated by parts, used the Fourier decompositions
(\ref{FTETAPHI}) and imposed an impenetrable bottom condition at
$z=-h$.  The wave energy density carried by wavevector $\k$ is
(\cite{GMMP})
\begin{equation}
{\cal E}(\k,t) 
={1\over 2}\mbox{Tr}\left[{\bf A}(\k)\W(\k,t)\right],
\label{EAW}
\end{equation}
where $A_{11}(\k)=k^{2}+1, A_{22}(\k)=k\tanh kh, A_{12}=A_{21}=0$.
Thus, the  Wigner distribution epitomises the local surface wave energy
density.


In the presence of slowly varying drift, we identify ${\bf
  W}(\x,\k,t)$ as the {\it local} Wigner distribution at position $\x$
representing waves of wavevector $\k$.  The time evolution of its
Fourier transform ${\bf W}(\p,\k,t)$, can be derived by considering
time evolution of the vector field $\psib$ implied by the boundary
conditions (\ref{BC}):
\begin{equation}
\begin{array}{ll}
\displaystyle \dot{\psi}_{j}(\k,t)+iL_{j\ell}(\k)\psi_{\ell}(\k,t)  = 
&\displaystyle i\int_{\q} \U(\q)\cdot
(\k-\q\delta_{j2})\psi_{j}(\k-\q,t) \\[13pt]
&+i\sqrt{\e}\int_{\q}
\d\U(\q)\cdot(\k-\q\delta_{j2}/\e)\psi_{j}(\k-\q/\e,t),
\end{array}
\label{EQMP}
\end{equation}
where the operator $\L(\k)$ is defined by
\begin{equation}
\L(\k) = \left(\begin{array}{cc} 
0 & i\vert\k\vert \tanh \e\vert\k\vert h \\[13pt]
-i(\e k^{2} +\e^{-1}) & 0 \end{array}\right).
\end{equation}
We have redefined the physical wavenumber to be $k/\e$ so that $k\sim
O(1)$.  Upon using (\ref{EQMP}) and the definition (\ref{WIGNERP}),
(see Appendix A)
\begin{equation}
\begin{array}{rl}
\displaystyle \dot{W}_{ij}(\p,\k,t) 
& \displaystyle = i W_{i\l}(\p,\k,t)
L_{\l j}^{\dagger}\left({\k\over \e}+
{\p\over 2}\right) - iL_{i\l}
\left({\k\over \e}-{\p\over 2}\right)
W_{\l j}(\p,\k,t) \\[13pt]
\: & \displaystyle 
+i\int_{\q} \U(\q)\cdot
\left(-{\k \over \e}+{\p \over 2}-\q\delta_{i2}\right) 
W_{ij}(\p-\q, \k+\e\q/2,t) \\[13pt]
\: & \displaystyle -i\int_{\q} 
\U(\q)\cdot\left(-{\k\over \e}-{\p\over
2}+\q\delta_{j2}\right)W_{ij}(\p-\q, \k-\e\q/2,t) \\[13pt]
\: & \displaystyle +i\sqrt{\e}\int_{\q} 
\d\U(\q)\cdot\left(-{\k \over \e}
+{\p \over 2}-{\q\over \e}\delta_{i2}\right) 
W_{ij}(\p-\q/\e,\k+\q/2,t) \\[13pt]
\: & \displaystyle -i\sqrt{\e}\int_{\q} \d\U(\q)\cdot
\left(-{\k\over \e}-{\p\over
2}+{\q\over \e}\delta_{j2}\right)W_{ij}(\p-\q/\e, \k-\q/2,t),
\end{array}
\label{WDOT2}
\end{equation}
where only the index $\ell = 1,2$ has been summed over.  If we now
assume that ${\bf W}(\x,\k,t)$ can be expanded in functions that vary
independently at the two relevant length scales, functions of the
field $\p$ (dual to $\x$) can be replaced by functions of a slow
variation in $\p$ and a fast oscillation $\xib/\e$; $\p\rightarrow\p
+\xib/\e$.


This amounts to the Fourier equivalent of a two-scale expansion in
which $\x$ is replaced by $\x$ and $\y=\x/\e$ (\cite{RPK}).  The two
new independent wavevectors $\p$ and $\xib$ are both of $O(1)$.
Expanding the Wigner distribution in powers of $\sqrt{\e}$ and using
$\p \rightarrow \p + \xib/\e$,
\begin{equation}
{\bf W}(\p,\k,t) \rightarrow {\bf W}_{0}(\p,\xib,\k,t)
+\sqrt{\e} {\bf W}_{1/2}(\p, \xib,\k,t) + \e {\bf W}_{1}
(\p, \xib,\k,t) + O(\e^{3/2}),
\end{equation}
we expand each quantity appearing in (\ref{WDOT2}) in powers
of $\sqrt{\e}$ and equate like powers.  Upon expanding the off-diagonal
operator $\L(-\k/\e+\p/2) = \e^{-1}{\bf L}_{0}(\k) +{\bf L}_{1}(\k,\p) +
O(\e)$, where
\begin{equation}
{\bf L}_{0}(\k)
= \left(\begin{array}{cc} 0 & ik\tanh kh \\[13pt]
-i(k^{2}+1) & 0  \end{array}\right),
\quad {\bf L}_{1}(\k,\p)  \equiv 
\left(\begin{array}{cc} 
\displaystyle 0 & i\p\cdot\k f(k) \\[13pt]
\displaystyle  i\p\cdot\k & 0
\end{array} \right)
\end{equation}
and 
\begin{equation}
f(k) \equiv -{hk +  \sinh kh \cosh kh \over 2k\cosh^{2}kh}.
\end{equation}

\subsection{Order $\e^{-1}$ terms}

The terms of $O(\e^{-1})$ in (\ref{WDOT2}) are
\begin{equation}
\begin{array}{l}
\displaystyle {\bf W}_{0}(\p,\xi,\k,t){\bf L}_{0}^{\dagger}
(\k_{+})-{\bf L}_{0}(\k_{-})
{\bf W}_{0}(\p,\xi,\k,t) = 0,\quad \,
\k_{\pm}\equiv \k\pm {\xib\over 2}
\end{array}
\label{ZERO}
\end{equation}
To solve (\ref{ZERO}), we use the eigenvalues and normalised
eigenvectors for ${\bf L}_{0}$ and its complex adjoint ${\bf
L}_{0}^{\dagger}$,
\begin{equation}
\displaystyle \tau\Omega(\k)-i\gamma,\,
{\bf b}_{\tau} =  \left(\begin{array}{c} \displaystyle 
i\tau\sqrt{\a(\k)/2} \\[13pt]
\displaystyle {1\over \sqrt{2\a(\k)}} \end{array}\right); \quad 
\displaystyle \tau\Omega(\k)+i\gamma,\, {\bf c}_{\tau} 
= \displaystyle  \left(\begin{array}{c} \displaystyle 
{i\tau\over \sqrt{2\a(\k)}} \\[16pt]
\displaystyle \sqrt{\a(\k)/2} \end{array}\right),
\label{EIGENB}
\end{equation}
where $\displaystyle \a(\k)\equiv {\Omega(\k)\over k^{2}+1}$, $\tau =
\pm 1$, and $i\gamma\rightarrow 0$ is a small imaginary term.  A ${\bf
  W}_{0}(\p,\xib,\k,t)$ that manifestly satisfies (\ref{EIGENB}) can
be constructed by expanding in the basis of $2\times 2$ matrices
composed from the eigenvectors:
\begin{equation}
\begin{array}{rl}
{\bf W}_{0}(\p,\xib,\k,t)&=\displaystyle
\sum_{\tau,\tau'=\pm}a_{\tau\tau'}(\p,\k,t)
{\bf b}_{\tau}(\k_{-}){\bf b}^{\dagger}_{\tau'}(\k_{+}).
\end{array}
\label{DECOMPW0}
\end{equation}
Right[left] multiplying (\ref{ZERO}) (using (\ref{DECOMPW0}))
by the eigenvectors of the adjoint problem, $\c_{\tau}(\k_{-})
\left[\c_{\tau}^{\dagger}(\k_{+})\right]$, we find that $a_{+-}=a_{-+} =
0$, and $a_{--}(\x,\k,t)\equiv a_{-}(\x,\k,t) =
a_{++}(\x,-\k,t)\equiv a_{+}(\x,-\k,t)$. Furthermore, $a_{+},a_{-}\neq 0$
only if $\xib = 0$.  Thus ${\bf W}_{0}$ has the form
\begin{equation}
{\bf W}_{0}(\p,\xib,\k,t) 
= {\bf W}_{0}(\p,\k,t)\d(\xib).
\end{equation}

From the definition of ${\bf W}_{0}$, we see that the (1,1) component of
${\bf W}_{0}$ is the local envelop of the ensemble averaged wave intensity
$\vert\eta(\x,\k,t)\vert^{2}\simeq a_{+}(\x,\k,t)\a(\k)$.  Similarly, from
the energy (Eq. (\ref{EAW})), we see immediately that the local ensemble
averaged energy density 
\begin{equation}
\label{eq:energy}
\begin{array}{rl}
\langle {\cal E}(\x,\k,t)\rangle &= A_{11}(\k)\a(\k)\langle a(\x,\k,t)\rangle 
+ A_{22}(\k)\langle a(\x,\k,t)\rangle  \\[13pt]
\: &= \Omega(\k)\langle a(\x,\k,t)\rangle.
\end{array}
\end{equation}
Therefore, since the starting dynamical equations are linear, we can
identify $\langle a(\x,\k,t)\rangle$ as the ensemble averaged local
wave action associated with waves of wavevector $\k$ (\cite{HENYEY}).
The wave action $\langle a(\x,\k,t)\rangle$, rather than the energy
density $\langle {\cal E}(\x,\k,t)\rangle$ is the conserved quantity
(\cite{LH,MEI,WHITHAM}).

The physical origin of $\gamma$ arises
from causality, but can also be explicitly derived from considerations
of an infinitesimally small viscous dissipation (\cite{CHOUB}). Although
we have assumed $\gamma \rightarrow 0$, for our model to be valid, the
viscosity need only be small enough such that surface waves are not
attenuated before they have a chance to multiply scatter and enter the
transport or diffusion regimes under consideration.  This constraint can
be quantified by noting that in the frequency domain, wave dissipation
is given by $\gamma = 2\nu k^{2}$ (\cite{LANDAU}) where $\nu$ is the
kinematic viscosity and 
\begin{equation}
  \label{eq:cg}
  c_{g}(k)\equiv \vert\nabla_{\k}\Omega(k)\vert
\end{equation}
is the group velocity.  The corresponding decay length $k_{d}^{-1}
\sim c_{g}(k)/(\nu k^{2})$ must be greater than the relevant wave
propagation distance. Therefore, we require
\begin{equation}
{\e^{2}c_{g}(k/\e)\over \nu k^{2}} 
\gg (1, \e^{-1})
\label{CRITERION1}
\end{equation}
for (transport, diffusion) theories
to be valid. The inequality (\ref{CRITERION1})
gives an upper bound for the viscosity
\begin{equation}
\nu k^{2} \ll (\e c_{g}(k/\e),
\e^{2}c_{g}(k/\e))
\end{equation}
which is most easily satisfied in the shallow water wave regime for
transport.  Otherwise we must at least require $\nu < o(\sqrt{\e})$.
The upper bounds for $\nu$ (and hence $\gamma$) given above provide
one criterion for the validity of transport theory.

\subsection{Order $\e^{-1/2}$ terms}

Collecting terms in (\ref{WDOT2}) of order $\e^{-1/2}$, we obtain
\begin{equation}
\begin{array}{l}
\displaystyle {\bf W}_{1/2}(\p,\xib,\k,t)
{\bf L}^{\dagger}_{0}(\k_{+})
-{\bf L}_{0}(\k_{-}){\bf W}_{1/2}(\p,\xib,\k,t) 
\displaystyle + \int_{\q} \U(\q)\cdot\xib {\bf
W}_{1/2}(\p-\q,\xib-\q,\k,t) \\[13pt]
\hspace{0.5in} \displaystyle 
- \int_{\q} \d\U(\q)\cdot\k_{-}\,{\bf W}_{0}
(\p,\xib-\q,\k+\q/2,t) +\int_{\q} \d\U(\q)\cdot\k_{+}\,
{\bf W}_{0}(\p,\xib-\q,\k-\q/2,t) \\[13pt]
\hspace{0.5in}
\displaystyle -\int_{\q} \d\U(\q)\cdot\q\left[
{\bf W}_{0}(\p,\xib-\q,\k+\q/2,t){\bf S}+
{\bf S}{\bf W}_{0}(\p,\xib-\q,\k-\q/2,t)\right] = 0
\end{array}
\label{SQRTE}
\end{equation}
where ${\bf S} = \left[\begin{array}{cc}
    0 & 0 \\
    0 & 1 \end{array}\right].$


Similarly decomposing ${\bf W}_{1/2}$ in the basis matrices composed
of ${\bf b}_{\tau}(\k_{-}){\bf b}^{\dagger}_{\tau'}(\k_{+})$ (as in
\ref{DECOMPW0}), substituting ${\bf W}_{0}(\p,0,\k,t)\d(\xib)$ from
(\ref{DECOMPW0} into (\ref{SQRTE}), and inverse Fourier transforming
in the slow variable $\p$, we obtain
\begin{equation}
\begin{array}{l}
\displaystyle {\bf W}_{1/2}(\x,\k,\xib,t) = \sum_{\tau,\tau'=\pm}
{\d\U(\xib)\cdot\Gb_{\tau,\tau'}(\x,\xib,\k,t) 
{\bf b}_{\tau}(\k_{-}){\bf b}^{\dagger}_{\tau'}(\k_{+}) \over 
\tau'\Omega(\k_{+})-
\tau\Omega(\k_{-})+\U(\x)\cdot\xib+2i\gamma},
\end{array}
\label{W1/2}
\end{equation}
where 
\begin{equation}
\begin{array}{l}
\Gb_{\tau,\tau'}(\x,\xib,\k,t) \equiv
\k_{-}a_{\tau'}(\x,\k_{+},t)\c_{\tau}^{\dagger}(\k_{-})\b_{\tau'}(\k_{+})
-\k_{+}a_{\tau}(\x,\k_{-},t)\b_{\tau}^{\dagger}(\k_{-})
\c_{\tau'}(\k_{+})\\[13pt]
\:\displaystyle \hspace{1in}+{\xib\over 2}\sum_{\mu=\pm}
\left[a_{\mu}(\x,\k_{+},t)\c_{\tau}^{\dagger}(\k_{-})
\b_{\mu}(\k_{+}) +
a_{\mu}(\x,\k_{-},t)\b_{\mu}^{\dagger}(\k_{-})\c_{\tau'}(\k_{+})\right].
\end{array}
\end{equation}

\subsection{Order $\e^{0}$ terms}
 
The terms of order $\e^{0}$ in (\ref{WDOT2}) read
\begin{equation}
\begin{array}{l}
\displaystyle \dot{{\bf W}}_{0}(\p,\k,t) =  \displaystyle
i\W_{0}(\p,\k,t)\L_{1}^{\dagger}(-\p)
-i\L_{1}(\p)\W_{0}(\p,\k,t)-i\int_{\q} \k\cdot\U(\q)
\q\cdot\nabla_{\k}\W_{0}(\p-\q,\xib,\k,t) \\[13pt]
\:\hspace{5mm} \displaystyle +i\int_{\q} \U(\q)
\cdot\p\,\W_{0}(\p-\q,\xib,\k,t) -i\int_{\q} \U(\q)\cdot\q
\left[{\bf S}\W_{0}(\p-\q,\xib,\k,t)+\W_{0}(\p-\q,\xib,\k,t){\bf S}\right]\\[13pt]
\: \hspace{5mm} \displaystyle +i\int_\q
\delta\U(\q)\cdot\k_{+}\,\W_{1/2}(\p,\xib-\q,\k-\q/2,t)
-i\int_\q \delta\U(\q)\cdot\k_{-}\,\W_{1/2}(\p,\xib-\q,\k+\q/2,t) \\[13pt]
\:\hspace{5mm} \displaystyle -\int_{\q}\delta\U(\q)\cdot\q
\left[{\bf S}\W_{1/2}(\p,\xib-\q,\k+\q/2,t)+\W_{1/2}(\p,\xib-\q,\k-\q/2,t){\bf S}
\right] \\[13pt]
\:\hspace{5mm} \displaystyle + i \W_{1}\L_{0}^{\dagger} -i\L_{0}\W_{1}
+\int_{\q} \U(\q)\cdot\xib \W_{1}(\p-\q,\xib,\k,t).
\end{array}
\label{ORDER0}
\end{equation}

To obtain an equation for the statistical ensemble average $\langle
a_{+}(\x,\k,t)\rangle$, we multiply (\ref{ORDER0}) by
$\c_+^\dagger(\k)$ on the left and by $\c_+(\k)$ on the right and
substitute ${\bf W}_{1/2}$ from equation (\ref{W1/2}). We obtain a
closed equation for $a(\x,\k,t)\equiv \langle a_{+}(\x,\k,t)\rangle$
(we henceforth suppress the $\langle \ldots\rangle$ notation for
$a(\x,\k,t)$ and ${\cal E}(\x,\k,t)$) by truncating terms containing
${\bf W}_{1}$.  Clearly, from (\ref{EIGENB}), $\c_+^\dagger(\k)(i
\W_{1}\L_{0}^{\dagger}-i\L_{0}\W_{1} )\c_+(\k)=0$.  Furthermore, we
assume $\langle \xib \W_{1}(\p-\q,\xib,\k,t) \rangle \approx 0$ which
follows from ergodicity of dynamical systems, and has been used in the
propagation of waves in random media (see \cite{RPK,BFPR}). The
transport equations resulting from this truncation are rigorously
justified in the scalar case (\cite{SPOHN,EY}).

\section{The surface wave transport equation}

The main mathematical result of this paper, an evolution equation for
the ensemble averaged wave action $a(\x,\k,t)$ follows from 
equation (\ref{ORDER0})
above (cf. Appendix A) and reads,
\begin{equation} 
\begin{array}{l} 
\dot{a}(\x,\k,t) +
\nabla_{\k}\omega(\x,\k)\cdot\nabla_{\x}a(\x,\k,t)
-\nabla_{\x}\omega(\x,\k)\cdot\nabla_{\k}a(\x,\k,t) \\[13pt] 
\hspace{3cm} \displaystyle = -\Sigma(\k)a(\x,\k,t) +
\int_{\q}\s(\q,\k)a(\x,\q,t),\gbout{\\[13pt] 
\displaystyle \hspace{3cm} + \,\mbox{terms proportional to}\,\,
R_{ij}(\q)q_{i}}
\end{array} 
\label{RESULT1} 
\end{equation} 
where 
\begin{equation}
\label{eq:omega} 
\omega(\x,\k) = \sqrt{(k^3+k)\tanh kh} + \U(\x)\cdot\k
\equiv \Omega(\k)+\U(\x)\cdot\k.  
\end{equation} 
The left hand side in (\ref{RESULT1}) corresponds to wave action
propagation in the absence of random fluctuations. It is equivalent to
the equations obtained by the ray theory, or a WKB expansion (see
section 5.1).  The two terms on the right hand side of (\ref{RESULT1})
represent refraction, or ``scattering'' of wave action out of and into
waves with wavevector $\k$ respectively. In deriving (\ref{RESULT1})
we have inverse Fourier transformed back to the slow field point
variable $\x$, and used the relation $(\a(\k)-f(k)\a^{-1}(\k))\k
\equiv \nabla_{\k}\Omega(\k)$.  To obtain (\ref{RESULT1}), we assumed
$R_{ij}(\q)q_{i}=R_{ij}(\q)q_{j}=0$, which would always be valid for
divergence-free flows in two dimensions.  Although the perturbation
$\delta \U$ is not divergence-free in general, ${\nobreak
  \nabla\!\cdot\!\delta \U(\x,z=0)} =-\partial_{z}\d U_{z}(\x,0) \ne
0$, using symmetry considerations, we will show in section
\ref{sub:rij} that $R_{ij}(\q)q_{i}=R_{ij}(\q)q_{j}=0$.

Explicitly, the scattering rates are
\begin{equation}
\begin{array}{rl}
\Sigma(\k) \equiv & \displaystyle 
2\pi \int_{\q}
q_{i}R_{ij}(\q-\k)k_{j}\sum_{\tau=\pm}
\b_{+}^{\dagger}(\k)\c_{\tau}(\q)\b_{\tau}^{\dagger}(\q)\c_{+}(\k) 
\d\left(\tau\omega(\x,\tau\q)-\omega(\x,\k)\right) \\[13pt]
\sigma(\q,\k)\equiv & \displaystyle
2\pi\sum_{\tau=\pm}
\tau q_{i}R_{ij}(\tau\q-\k)k_{j}
\vert\b_{\tau}^{\dagger}(\tau\q)
\c_{+}(\k)\vert^{2}\d
\left(\tau\omega(\x,\q)-\omega(\x,\k)\right)
\end{array}
\label{SECTIONS}
\end{equation}
where 
\begin{equation}
\begin{array}{rl}
\b_{+}^{\dagger}(\k)\c_{\tau}(\q)\b_{\tau}^{\dagger}(\q)\c_{+}(\k) & 
=\displaystyle 
{(\tau\a(\k)+\a(\q))(\tau\a(\q)+\a(\k))\over 4\a(\k)\a(\q)} \\[13pt]
\vert\b_{\tau}^{\dagger}(\k)\c_{\tau'}(\q)
\vert^{2} & = \displaystyle
{(\tau\a(\q)+\a(\k))^{2}\over 4\a(\k)\a(\q)}.
\end{array}
\end{equation}
Physically, $\Sigma(\k)$ is a decay rate arising from scattering of
action out of wavevector $\k$. The kernel $\sigma(\q,\k)$ represents
scattering of action from wavevector $\q$ {\it into} action with
wavevector $\k$.  Note that the slowly varying drift $\U(\x)$ enters
parametrically in the scattering via $\omega(\x,\k)$ in the
$\delta-$function supports.  The arguments $\omega(\x,\k)$ in the
$\delta-$functions mean that we can consider the transport of waves of
each fixed frequency $\omega_{0}\equiv\omega(\x,\k)$ independently.

The typical distance travelled by a wave before it is significantly
redirected is defined by the mean free path
\begin{equation}
\ell_{\mbox{{\it mfp}}} = {c_{g}(k) \over \Sigma(k)} \sim O(1).
\label{MFP}
\end{equation}
The mean free path described here carries a different interpretation
from that considered in weakly nonlinear, or multiple scattering
theories (\cite{ZAK}) where one treats a low density of scatterers.
Rather than strong, rare scatterings over every distance
$\ell_{\mbox{{\it mfp}}}\sim O(1)$, we have considered constant, but
weak interaction with an extended, random flow field. Although here,
each scattering is $O(\e)$ and weak, over a distance of $O(1)$,
approximately $\e^{-1}$ interactions arise, ultimately producing
$\ell_{\mbox{{\it mfp}}}\sim O(1)$.



\section{Results and Discussion}

We have derived transport equations for water wave propagation
interacting with static, random surface flows containing two explicit
length scales. We have further assumed that the amplitude of $\d\U$
scales as $\e^{\beta}$ with $\beta = 1/2$: The random flows are
correspondingly weakened as the high frequency limit is taken.  Since
scattering strength is proportional to the power spectrum of the
random flows and is quadratic in $\delta \U$, the mean free path can
be estimated heuristically by $\ell_{\mbox{{\it mfp}}} \sim
c_{g}(k)/\Sigma(k) \e^{1-2\beta}$.  For $\beta > 1/2$, the scattering
is too weak and the mean free path diverges.  In this limit, waves are
nearly freely propagating and can be described by the slowly varying
flows alone, or WKB theory. If $\beta < 1/2$, $\ell_{\mbox{{\it mfp}}}
\rightarrow 0$ and the scattering becomes so frequent that over a
propagation distance of $O(1)$, the large number of scatterings lead
to diffusive (cf. Section 5.4) behaviour (\cite{SHENG}).  Therefore,
only random flows that have the scaling $\beta = 1/2$ contribute to
the wave transport regime.

We also note that $\beta > 0$ precludes any wave localisation
phenomena.  In a two-dimensional random environment, the localisation
length over which wave diffusion is inhibited is approximately
(\cite{SHENG})
 \begin{equation}
\ell_{\mbox{{\it loc}}} \sim \ell_{\mbox{{\it mfp}}} \exp\left(\e^{-1}k
\ell_{\mbox{{\it mfp}}}\right) \sim
\e^{1-2\beta}\exp\left(\e^{-2\beta}\right).
\end{equation}
As long as the random potential is scaled weaker $(\beta > 0)$,
$\ell_{\mbox{{\it loc}}} \rightarrow \infty$, and strong localisation
will not take hold. In the following subsections, we systematically
discuss the salient features of water wave transport contained in Eq.
(\ref{RESULT1}) and derive wave diffusion for propagation distances
$\gtrsim O(1)$.

\subsection{Slowly varying drift: $\U(\x)\neq 0, \d\U=0$}

First consider the case where surface flows vary only on
scales much larger than the longest wavelength $2\pi/k$
considered, {\it i.e.,} $\d\U = 0$.  The left-hand side in
(\ref{RESULT1}) represents wave action transport over slowly
varying drift and may describe short wavelength modes propagating
over flows generated by underlying long ocean waves.  

We first demonstrate that the nonscattering terms of the transport
equation (\ref{RESULT1}) is equivalent to the results obtained by ray
theory (WKB expansion) and conservation of wave action (CWA)
(\cite{LH,MEI,PEREGRINE76,WHITE,WHITHAM}).  Assume the WKB expansion
(\cite{KELLER,BENDER})
\begin{equation}
  \eta_\e=A_\eta(\x,t) e^{iS(\x,t)/\e} \quad \mbox{ and } \quad
  \v_\e=A_\v(\x,t) e^{iS(\x,t)/\e},
\end{equation}
with smoothly varying $A_\eta$ and $A_\v$.  Upon using the above {\it
  ansatz} in (\ref{eq:wigner}) and setting $\e\to0$, we have
$a(\x,\k,t)=|A|^2(\x,t)\delta(\k-\nabla_\x S(\x,t))$ where
$|A|^2=2\alpha(k)|A_\v|^2=2\alpha^{-1}(k)|A_\eta|^2$. Substitution of
this expression for $a(\x,\k,t)$ into (\ref{RESULT1}), we obtain the
following possible equations for $S(\x,t)$ and $|A|^2(\x,t)$
\begin{eqnarray}
  \label{eq:eiconal}
  \partial_t S + \omega(\x,\nabla_\x S) &=& 0, \\
\label{eq:transsampl}
  \partial_t |A|^2(\x,t) +\nabla_{\x}\cdot
\Big( |A|^2 \nabla_\k\omega(\x,\nabla_\x S) \Big) &=& 0.
\end{eqnarray}
The first equation is the eikonal equation, while the second
equation is the wave action amplitude equation. Recalling that
$|A_\eta|^2=\alpha(k)|A|^2/2$, we obtain the following transport
equation for the height amplitude:
\begin{equation}
  \label{eq:transheight}
  \partial_t \Big(\frac{|A_\eta|^2}{\alpha(\nabla_\x S)}\Big) +\nabla_{\x}\cdot 
\Big( \frac{|A_\eta|^2}{\alpha(\nabla_\x S)} 
   \nabla_\k\omega(\x,\nabla_\x S) \Big) = 0.
\end{equation}
Equation (\ref{eq:transheight}) is the same as Eq. (8) of
\cite{WHITE}, except that his $\bar\Omega$ is replaced here with
$\alpha$ due to our inclusion of surface tension.

Wave action conservation can be understood by noting that
\begin{equation}
  \label{eq:consact}
  \frac{d}{dt} a(X(t),K(t),t)=0,
\end{equation}
where the characteristics $(X(t),K(t))$ satisfy the Hamilton equations
\begin{equation}
  \label{eq:Ham}
  \frac{dX(t)}{dt}=\nabla_\k\omega(X(t),K(t)), \qquad \mbox{ and } \qquad
  \frac{dK(t)}{dt}=-\nabla_\x\omega(X(t),K(t)).
\end{equation}
The solutions to the ordinary differential equations
(\ref{eq:Ham}) are the characteristic curves used to solve
(\ref{eq:eiconal}) and (\ref{eq:transsampl}) (\cite{Courant-Hilbert}).

\subsection{Correlation functions and conservation laws}
\label{sub:rij}

We now consider the case where $\delta\U\ne0$.  The scattering rates
defined by (\ref{SECTIONS}) depend upon the precise form of the random
flow correlation $R_{ij}$. There are actually six additional terms in
(\ref{SECTIONS}) in the calculation of $\sigma$ and $\Sigma$, which
vanish because
\begin{equation}
  \label{eq:rijzero}
  \sum_{j=1}^2 R_{ij}(\q)q_j=0 \qquad \mbox{ for } i=1,2.
\end{equation}
We prove relation (\ref{eq:rijzero}) provided that $\delta
U_z(\k,k_z)$ and $\delta U_z(\k,-k_z)$ have the same probability
distribution. Thus,
\begin{equation}
  \label{eq:symprob}
\langle\delta U_i(\p,p_z) \delta U_z(\k,k_z)\rangle 
= \langle\delta U_i(\p,p_z) \delta U_z(\k,-k_z)\rangle
\end{equation}
This symmetry condition is reasonable, and is compatible with the
divergence-free condition for $\delta \U$ in three dimensions.  We
show that Hypothesis (\ref{eq:symprob}) implies (\ref{eq:rijzero}) by
first using incompressibility $\sum_{j=1}^2 \delta
U_j(\k,k_z)k_j+\delta U_z(\k,k_z)k_z=0$:
\begin{displaymath}
  \begin{array}{rcl}
\displaystyle
\sum_{j=1}^2 \delta(\p+\k) R_{ij}(\k) k_j &=& 
\displaystyle
\sum_{j=1}^2 \langle \delta U_i(\p,0)\delta U_j(\k,0) k_j \rangle \\
 &=& 
\displaystyle
\sum_{j=1}^2 \langle \delta U_i(\p,0)\displaystyle\int_{k_z}
   \delta U_j(\k,k_z) k_j  \rangle \\
 &=& 
- \displaystyle\int_{-\infty}^\infty
    \langle \delta U_i(\p,0)
   \delta U_z(\k,k_z) \rangle k_z dk_z \\
&=& 0, 
  \end{array}
\end{displaymath}
where the last equality follows from (\ref{eq:symprob}). Thus,
(\ref{eq:rijzero}) is verified, and (\ref{SECTIONS}) derived.

The form $R_{ij}(\vert\q\vert)q_{i} = R_{ij}(\vert\q\vert)q_{j} = 0$,
requires the correlation function to be transverse:
\begin{equation}
R_{ij}(\vert\q\vert) = R(q)\left[\d_{ij} - {q_{i}q_{j}\over q^{2}}\right],
\label{TRANSVERSE}
\end{equation}
where $R(q)$ is a scalar function of $q$.  The correlation kernels in
the scattering integrals can now be written as
\begin{equation}
\begin{array}{rl}
q_{i}R_{ij}(\vert\tau \q-\k\vert)k_{j} & \displaystyle
= R(\vert\tau\q-\k\vert)\left[\q\cdot\k-
{\q\cdot(\tau\q-\k)\k\cdot(\tau\q-\k)\over \vert\tau\q-\k\vert^{2}}\right] \\[13pt]
\: & \displaystyle = \tau {R(\vert\tau\q-\k\vert)\over \vert\tau\q-\k\vert^{2}}
q^{2}k^{2}\sin^{2}\theta
\end{array}
\label{TRANSVERSE2}
\end{equation}
where $\theta$ denotes the angle between $\q$ and $\k$.  The
scattering must also satisfy the support of the $\d$-functions; for
$\U(\x)=0$ only $\vert\q\vert = \vert\k\vert$ satisfy the the
$\delta-$function constraints. In the presence of slowly varying
drift, the evolution of $a(\x,\vert\k\vert \neq \vert\q\vert)$ can
``doppler'' couple to $a(\x,\q,t)$.

It is straightforward to show from the explicit expressions
(\ref{SECTIONS}) that
\begin{equation}
\label{eqconsscat}
    \Sigma(\k)=\int_{\q}\sigma(\k,\q).
\end{equation}
This relation indicates that the scattering operator on the right hand
side of (\ref{RESULT1}) is conservative: Integrating (\ref{RESULT1})
over the whole phase space yields
\begin{equation}
  \label{eq:conswaveaction}
  \dfrac{d}{dt}  \dint_\x\dint_\k a(\x,\k,t) \,=\, 0.
\end{equation}
Equation (\ref{eq:conswaveaction}) is the generalization of CWA to
include scattering of action from rapidly varying random flows
$\d\U(\x/\e)$.  Although $a(\x,\k,t)$ is conserved, the total water
wave energy ${\cal E}(\x,\k,t)=\Omega(\k)a(\x,\k,t)$ will not be
conserved.  For example, if $U(\x)$ is small enough such that the
$\d-$function in the $\sigma(\q,\k)$ integral is triggered only when
$\tau=+1$,
\begin{equation}
  \label{eq:nonconsen}
  \dfrac{d}{dt} {\cal E} = \dfrac{d}{dt}\dint_\x\dint_\k 
[\omega(\x,\k)-\k\cdot\U(\x)]
a(\x,\k,t) = - \dfrac{d}{dt}\dint_\x\dint_\k \k\cdot\U(\x) \,a(\x,\k,t)\neq 0.
\end{equation}
This nonconservation results from the energy that must be supplied in
order to sustain the stationary underlying flow. For small $U(\x)$,
the quantity $\omega(\x,\k)a(\x,\k,t)$ is conserved. In that case, the
evolution of $\omega(\x,\k)a(\x,\k,t)$ obeys an equation identical to
(\ref{RESULT1}).  When there is doppler coupling with $\tau=-1$, an
additional term arises and $\omega(\x,\k)a(\x,\k,t)$ is no longer
conserved under scattering.

\subsection{Doppler coupled scattering}

In addition to the correlation functions, the wave action scattering
terms involving $\Sigma(\k)$ and integrals over $\sigma(\q,\k)$ depend
also on the support of the $\delta-$function.  Consider action
contained in water waves of fixed wavevector $\k$.  When $\U(\x)=0$,
only $\tau=+1$ terms contribute to the the integration over $\q$ as
long as $\vert\q\vert = \vert\k\vert$.  In this case, we can define
the angle $\q\cdot\k=k^{2}\cos\theta$ and reduce the cross-sections to
single angular integrals over
\begin{equation}
\label{eq:Rijsimple}
q_i R_{ij}(|\q-\k|)q_j =
R\left(\big| 2k\sin{\theta\over 2}\big|\right) {k^{2}\over 4} 
{\sin^{2}\theta \over \sin^{2}{\theta\over 2}}, \quad
\tau=+1.
\end{equation}
In this case ($\U(\x)=0$), assuming $R(\vert\q\vert)$
is monotonically decreasing, the most important contribution to the
scattering occurs when $\q$ and $\k$ are collinear.

\begin{center}
\begin{figure}
  \epsfig{file=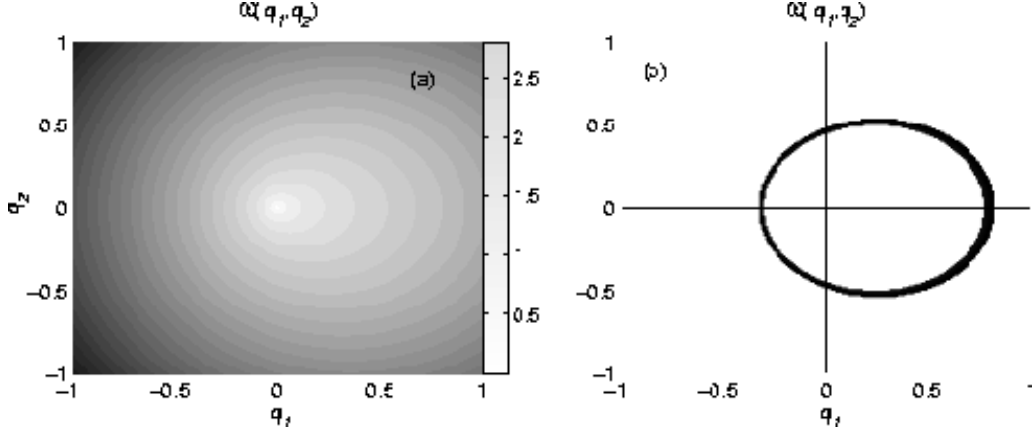,width = 13.6cm}
\caption{(a). Contour plot of $\omega(\q)$. Each grayscale corresponds to 
a different constant value of $\omega(\q)=\omega(\k)\equiv \omega_{0}$.
(b). The band of $\q$ that satisfies $0.625 < \omega_{0} < 0.6625$.
Wavevectors $\q$ and $\k$ that lie in this band can couple
$a(\x,\k,t)$ to $a(\x,\q,t)$ via wave scattering.}
\label{FIG3}
\end{figure} 
\end{center}

When $\U(\x)\neq 0$, and $\tau=+1$, the sets of $\q$ which satisfy
$\Omega(\q)+\U(\x)\cdot\q = \Omega(\k)+\U(\x)\cdot\k \equiv
\omega_{0}$ trace out closed ellipse-like curves and are shown in the
contour plots of $\omega(\q)$ in Figure \ref{FIG3}(a).  The parameters
used are $\U(\x)\cdot\k_{1} = -0.5 k_{1}$ and $h=\infty$ (the
$-\k_{1}, -\q_{1}$ directions are defined by the direction of
$\U(\x)$).
Each grayscale corresponds to a curve defined by fixed
$\omega(\k)=\omega_{0}$. All wavevectors $\q$ in each contour
contribute to the integration in the expressions for $\Sigma(\k)$ and
$\omega(\q,\k)$.  Thus, slowly varying drift can induce an indirect
doppler coupling between waves with different wavenumbers, with the
most drastic coupling occurring at the two far ends of a particular
oval curve.  For example, in Figure \ref{FIG3}(b), the dark band
denotes $\q$ such that $\omega(\q) = \omega_{0}$ when $0.625 <
\omega_{0} < 0.6625$.  The wavevectors $\q \approx (-0.3,0)$ and $\q
\approx (0.8,0)$ are two of many that contribute to the scattering
terms.  Therefore, the evolution of $a(\x,\k\approx (-0.3,0),t)$ also
depends on $a(\x,\q \approx (0.8,0),t)$ via the second term on the
right side of (\ref{RESULT1}).

\begin{center}
\begin{figure}
  \epsfig{file=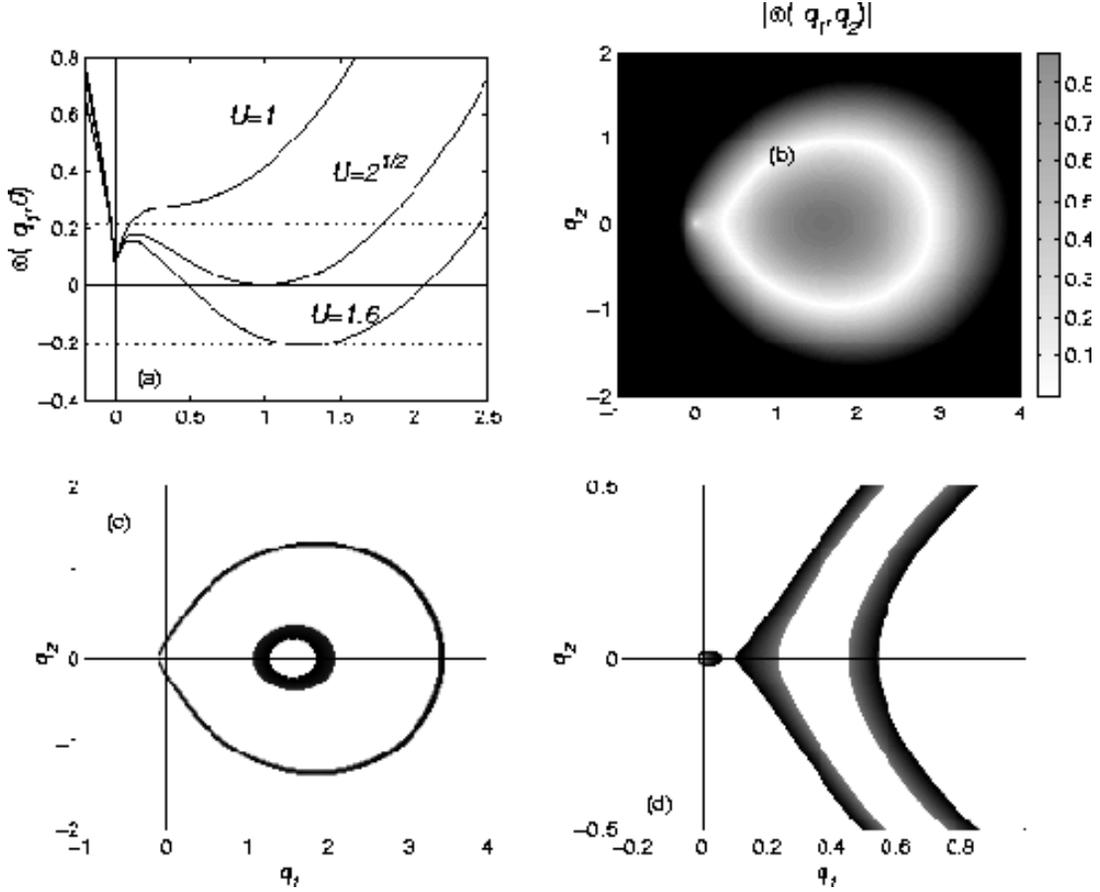,width = 14.4cm}
\caption{Conditions for doppler coupling when $\tau=-1$. 
(a). Plot of $\omega(q_{1},q_{2}=0; h=\infty)$ for $U=1$, $U=\sqrt{2}$, and
$U=1.6$. Only for $U>\sqrt{2}$ does $\omega(q_{1},q_{2}=0;h=\infty)<0$. 
(b). Contour plot of $\vert\omega(\q)\vert$. Each grayscale corresponds to 
a different constant value of $\omega(\q)=\omega(\k)\equiv -\omega_{0}$.
(c). The bands of $\q$ satisfying $0.414 < -\omega_{0} < 0.468$.
(d). An expanded view of the coupling bands for $0.0756<-\omega_{0}< 0.1368$.
Note that wavenumbers of very
small modulus can couple with wavenumbers of significantly larger
modulus.}
\label{FIG4}
\end{figure} 
\end{center}

Provided $\U(\x)$ is sufficiently large, the $\tau = -1$ terms can
also contribute to scattering.  The dissipative scattering rate
$\Sigma(\k)a(\x,\k,t)$ will change quantitatively since additional
$\q$'s will contribute to $\Sigma(\k)$.  However, this decay process
depends only on $\k$ and is not coupled to
$a(\x,\vert\q\vert\neq\vert\k\vert,t)$.  Wavevectors $\q$ that satisfy
the $\delta-$function in the $\sigma(\q,\k)a(\x,\q,t)$ term will, as
when $\tau=+1$, lead to indirect doppler coupling. This occurs when
$\omega(\q) = -\omega_{0}$ and, as we shall see, allows doppler
coupling of waves with more widely varying wavelengths than compared
to the $\tau=+1$ case. Observe that if $\tau=-1$ terms arise, the
drift frame energy $a(\x,\k,t)\omega(\x,\k)$ is no longer conserved.
Figure \ref{FIG4}(a) plots $\omega(q_{1}, q_{2}=0)$ for
$U(\x)=1<\sqrt{2}, U(\x)=\sqrt{2}$, and $U(\x)=1.6 >\sqrt{2}$.  Since
$\omega_{0}$ and $\omega(\q)$ are identical functions, $-\omega_{0}$
can take on values below the upper dotted line ($\omega_{0}\lesssim
0.22$ for $U=1.6$).  Therefore, coupling for $\tau=-1$ and $q_{2}=0$
occurs for values of $-\omega_{0}$ between the dotted lines.  Note
that depending upon the value of $\omega_{0}$, coupling can occur at
two or four different points $\q=(q_{1},0)$.  Figure \ref{FIG4}(b)
shows a contour plot of $\vert\omega(\q)\vert$ as a function of
$(q_{1},q_{2})$. A level set lying between the dotted lines in $(a)$
will slice out two bands; one band corresponds to all values of $\k$
that couple to $\q$ lying in the associated second band.  The two
bands determined by the interval $0.414 < -\omega_{0} < 0.468$ are
shown in Fig.  \ref{FIG4}(c). For any $\k$ lying in the inner band of
Fig.  \ref{FIG4}(c), all $\q$ lying in the outer band will contribute
to doppler coupling for $\tau=-1$, and {\it vice versa}. As
$-\omega_{0}$ is increased, the inner(outer) band decreases(increases)
in size, with the central band vanishing when $-\omega_{0}$ approaches
the upper dotted line in $(a)$ where the $\tau=-1$ coupling
evaporates. If $-\omega_{0}$ is decreased, the two bands merge, then
disappear as $-\omega_{0}$ reaches the lower limit.  Fig.
\ref{FIG4}(d) is an expanded view of the two bands for small
$0.0756<-\omega_{0}< 0.1368$.  Note that a small island of $\q$ or
$\k$ appears for very small wavevectors. The water wave scattering
represented by $\sigma(\q,\k)$ can therefore couple very long
wavelength modes with very short wavelength modes (the two larger
bands to the right in Fig.  \ref{FIG4}(d)).  However, the strength of
this coupling is still determined by the magnitude of
$q_{i}R_{ij}(\vert\q-\k\vert)k_{j}$, which may be small for
$\vert\q-\k\vert$ large.

The depth dependence of doppler coupling will be relevant when $hq, hk
\lesssim 1$ where $q$ and $k$ are the magnitudes of the wavevectors of
two doppler-coupled waves.  For $\tau=+1$, finite depth reduces the
ellipticity of the coupling bands, resulting in weaker doppler
effects.  Since the water wave phase velocity decreases with $h$, a
finite depth will also reduce the critical $U(\x)$ required for
$\tau=-1$ doppler coupling.  For small $U(\x)$, it is clear that the
$\d-$functions associated with the $\tau=-1$ terms in $\sigma(\q,\k)$
are first triggered when the $\q$ and $\k$ are antiparallel,
$\U\cdot\k = -k\vert\U\vert, \U\cdot\q = +q\vert\U\vert$.

\begin{center}
\begin{figure}
  \epsfig{file=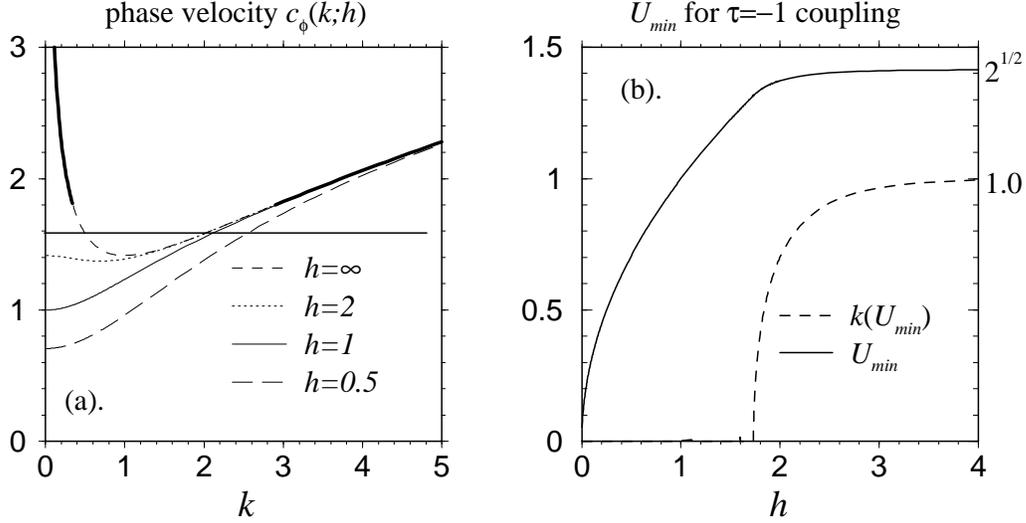,width = 13.6cm}
\caption{$U> c_{\phi}(k)$ is required for $\tau=-1$ coupling. 
(a). The phase velocity $c_{\phi}(k)$ for various depths $h$.
The velocity shown by the solid horizontal line $U\approx 1.6 > 
c_{\phi}(k; h=\infty)$ for $0.5 \lesssim k \lesssim 2$.
(b). The minimum $U_{min}(h)$ required for 
existence of $\tau=-1$ coupling at any
wavevector $k$, and the wavevector $k(U_{min})$ at which this first happens.}
\label{FIG5}
\end{figure} 
\end{center}

Figure \ref{FIG5}(a) shows the phase velocity for various depths $h$.
In order for $\tau=-1$ to contribute to scattering, $U \geq
c_{\phi}(k;h)$. For $U \approx 1.6$, this condition holds in the
$h=\infty$ case for $0.5 \lesssim k \lesssim 2$ (the dashed region of
$c_{\phi}(k,\infty)$). Recall that our starting equations (\ref{BC})
are valid only in the small Froude number limit.  However, for water
waves propagating over infinite depth, $\tau=-1$ coupling requires
$U>U_{min}= \mbox{min}_{k}\{c_{\phi}(k)\}$, with
$c_{\phi}(k_{min})\simeq 22$cm/s.  Therefore, in such ``supersonic''
cases, where $\tau=-1$ is relevant, our treatment is accurate only at
wavevectors $k^{*}$ such that $U \ll c_{\phi}(k^{*};h)$, {\it e.g.,}
the thick solid portion of $c_{\phi}(k;\infty)$.  For $U \gtrsim
U_{min}$, the $\tau=-1$ term can couple wavevectors $q \approx 0 \ll
k_{min}$ with $k \approx 2-3 \gg k_{min}$. The rich $\tau=-1$ doppler
coupling displayed in Figures \ref{FIG4} is particular to water waves
with a dispersion relation $\omega(\q)$ that behaves as $q^{3/2},
\U\cdot\q$, or $q^{1/2}$ depending on the wavelength.  Doppler
coupling in water wave propagation is very different from that arising
in acoustic wave propagation in an incompressible, randomly flowing
fluid (\cite{HOWE,RYZHIK,HUNTER}) where $\omega(\q) =
c_{s}\vert\q\vert$.  An additional doppler coupling analogous to the
$\tau=-1$ coupling for water waves arises only for supersonic random
flows when $U(\x) \geq c_{s}$, independent of $q$. In such instances,
compressibility effects must also be considered.

Figure \ref{FIG5}(b) plots the minimum drift velocity $U_{min}(h)$
where $\tau = -1$ doppler coupling first occurs at any wavevector. The
wavevector at which coupling first occurs is also shown by the dashed
curve. For shallow water, $h \ll\sqrt{3}$, $U_{min}(h) \propto
\sqrt{h}$ and very long wavelengths couple first (small $k(U_{min})$).
For depths $h > \sqrt{3}$ ($\sim 3$cm for water), the minimum drift
required quickly increases to $U^{*}(\infty)=\sqrt{2}$, while the
initial coupling occurs at increasing wavevectors until at infinite
depth, where the first wavevector to doppler couple approaches
$k\rightarrow 1$ (in water, this corresponds to wavelengths of $\sim
6.3$cm).  The conditions for $\tau = -1$ doppler coupling outlined in
Figures \ref{FIG3} and \ref{FIG4} apply to both $\Sigma(\k)$ and
$\sigma(\q,\k)$, with the proviso that $\q$ and $\k$ are parallel for
$\Sigma(\k)$ and antiparallel for $\sigma(\q,\k)$.  However, even when
$U<U_{min}$ such that only $\tau=+1$ applies, the set of $\q$
corresponding to a constant value of $\omega(\k)=\omega_{0}$, traces
out a noncircular curve.  There is doppler coupling between
wavenumbers $q\neq k$ as long as $U\neq 0$.

\subsection{Surface wave diffusion}

We now consider the radiative transfer equation (\ref{RESULT1}) over
propagation distances long compared to the mean free path
$\ell_{\mbox{{\it mfp}}}$.  Imposing an additional rescaling and
measuring all distances in terms of the mean free path, we introduce
another scaling $\epsilon^{-1}$, proportional to the number of mean
free paths travelled.  Since $\beta = 1/2$, transport of wave action
prevails when $O(\e) < \vert\x\vert \sim O(1)$, while diffusion holds
when $O(\epsilon^{-1}) \sim \vert\x\vert < \ell_{\mbox{\it loc}}$.

Since waves of each frequency satisfy (\ref{RESULT1}) independently,
we consider the diffusion of waves of constant frequency $\omega_{0}$.
To derive the diffusion equation, we assume for simplicity that $\U$
is constant and small such that $\omega_{0}+\omega(\x,\q)\neq0$ (the
$\tau=-1$ terms are never triggered by the $\delta-$functions).
Expanding all quantities in the transport equation (\ref{RESULT1}) in
powers of $\epsilon$, we find
\begin{equation} 
\label{eq:diffa0} 
\dot{a}_0 +\bar\U \cdot\nabla_\x a_0
- \nabla_\x \cdot {\bf D} \cdot \nabla_\x a_0 = 0.  
\end{equation}
The derivation of this equation is given in Appendix B.  The diffusion
tensor ${\bf D}$ is given in (\ref{eq:coeffD}) and is a function of
the power spectrum $R_{ij}$.  The effective drift $\bar\U$ is given by
(\ref{eq:barU}):
\begin{equation}
  \label{eq:barU2}
  \bar\U=\frac{\int_\k \nabla_\k \omega(\k)
         \delta(\k\cdot\U+\Omega(\k)-\omega_{0})}
    {\int_\k \delta(\k\cdot\U+\Omega(\k)-\omega_{0})}.
\end{equation}
Up to a change of basis, we can assume that $\U=U{\bf e}_1$, where
$U>0$.  Then the set of points $\k\cdot\U+\Omega(\k)-\omega_{0}=0$ is
symmetric with respect to the $x_{1}-$axis and $\bar\U$ is parallel to
$\U$. Also notice that the total energy given in (\ref{eq:energy}) is
asymptotically conserved in the diffusive regime.  Indeed, the total
energy variations are given by (\ref{eq:nonconsen}).  Assuming that
all water waves have frequency $\omega_0$, we have in the diffusive
regime
\begin{displaymath}
  \begin{array}{rcl}
\dfrac{d}{dt} {\cal E} &=&  
 - \dfrac{d}{dt}\dint_\x\dint_\k \k\cdot\U(\x) \,a(\x,\k,t) \\
 &\approx& -\Big(\dint_\x \dot{a}_0(\x,t)\Big) 
      \dint_\k \k\cdot\U \delta(\omega_0-\omega(\k)), \\
  \end{array}
\end{displaymath}
since $\U$ is constant. Recasting the diffusion equation as
$\dot{a}_0=-\nabla_\x\cdot(\bar\U a_0 + {\bf D} \cdot \nabla_\x a_0)$,
we deduce that
\begin{displaymath}
  \dint_\x \dot{a}_0(\x,t) =0,
\end{displaymath}
which conserves the total energy ${\cal E}$.

Now consider the simplified case $\U\equiv0$, $h=\infty$ and
$\Omega_\infty(\k)=\sqrt{k^3+k}$. Since $\U=\bar{\U}=0$,
(\ref{eq:Rijsimple}) holds and we have for all $\k$,

\begin{equation}
\label{eq:rijmean}
  \int_\q q_i R_{ij}(|\q-\k|)q_j \q = \bf{0}.
\end{equation}
We deduce that the corrector $\bchi$ in (\ref{eq:chi}) is given by

\begin{displaymath}
  \bchi(\k)=-\frac{\nabla_\k
\Omega_\infty(\k)}{\Sigma(k)}=-\frac{|\nabla_\k\Omega_\infty(\k)|}
{\Sigma(k)}\hat\k
=-\frac{c_g}{\Sigma(k)}\hat\k,
\end{displaymath}
where $\k = k\hat{\k}$. The isotropic diffusion tensor ${\bf D}$ is
thus given by

\begin{equation}
  \label{eq:Dsimple}
  {\bf D}=\frac{1}{\Sigma(k) V_{\omega_0}} 
   \int_\q |\nabla_\q\Omega_\infty(\q)|^2  \hat\q\hat\q^T
\delta(\Omega_\infty(\q)-\omega_0) = 
    \frac{c_g^2(k)}{2\Sigma(k)} \,\, {\bf I},
\end{equation}
where ${\bf I}$ is the $2\times2$ identity matrix. Thus, the diffusion
equation for $a_0(\x,t)$ assumes the standard form (\cite{SHENG})
\begin{equation}
  \label{eq:diffsimp}
        \dot{a}_0 - \frac{c_g^2(k)}{2\Sigma(k)}\Delta a_0 = 0.
\end{equation}

\section{Summary and Conclusions}

In this paper, we have used the Wigner distribution to derive the
transport equations for water wave propagation over a spatially random
drift composed of a slowly varying part $\U(\x)$, and a rapidly
varying part $\sqrt{\e}\d\U(\x/\e)$. The slowly varying part
determines the characteristics on which the waves propagate.  We
recover the standard result obtained from WKB theory: conservation of
wave action.  Provided $R_{ij}(\q)q_{j}=0$, we extend CWA to include
wave scattering from correlations $R_{ij}$ of the rapidly varying
random flow.  Evolution equations for the nonconserved wave intensity
and energy density can be readily obtained from (\ref{RESULT1}).
Moreover, conservation of drift frame energy $a(\x,\k,t)\omega(\x,\k)$
requires small $U<U_{min}$ and absence of $\tau=-1$ contributions to
scattering.

Explicit expressions for the scattering rates $\Sigma(\k)$ and
$\sigma(\q,\k)$ are given in Eqs. (\ref{SECTIONS}). For fixed
$\omega(\k)$, we find the set of $\q$ such that the $\d-$functions in
(\ref{SECTIONS}) are supported. This set of $\q$ indicates the
wavevectors of the background surface flow that can mediate doppler
coupling of the water waves. Although widely varying wavenumbers can
doppler couple, supported by the $\delta-$function constraints,
particularly for $\tau=-1$, the correlation $R_{ij}(\vert\q-\k\vert)$
also decreases for large $\vert\q-\k\vert$. For long times, multiple
weak scattering nonetheless exchanges action among disparate
wavenumbers within the transport regime. Our collective results,
including water wave action diffusion, provide a model for describing
linear ocean wave propagation over random flows of different length
scales. The scattering terms in (\ref{RESULT1}) also provide a means
to correlate sea surface wave spectra to statistics $R_{ij}$ of finer
scale random flows.

Although many situations arise where the underlying flow is rotational
(\cite{WHITE}), the irrotational approximation used simplifies the
treatment and allows a relatively simple derivation of the transport
and diffusive regimes of water wave propagation.  The recent extension
by \cite{WHITE} of CWA to include rotational flows also suggests that
an explicit consideration of velocity and pressure can be used to
generalise the present study to include rotational random flows. Other
feasible extensions include the analysis of a time varying random
flow, as well as separating the underlying flows into static and wave
dynamic components.

\acknowledgements

The authors thank A. Balk, M. Moscoso, G. Papanicolaou, L. Ryzhik, and
I. Smolyarenko for helpful comments and discussion. GB was supported
by AFOSR grant 49620-98-1-0211 and NSF grant DMS-9709320. TC was
supported by NSF grant DMS-9804780.

\appendix

\section{Derivation of the transport equation}

Some of the steps in the derivation of (\ref{RESULT1}) are outlined
here.  By taking the time derivative of $W_{ij}$ in (\ref{WIGNERP})
and using the definition (\ref{EQMP}) for $\dot{\psi}$, we obtain
\begin{equation}
\begin{array}{rl}
\displaystyle (2\pi\e)^{2}\dot{W}_{ij}(\p,\k,t) & \displaystyle =  (2\pi\e)^{2}i
W_{i\l}(\p,\k)L_{\l j}^{*}\left({\k\over \e}-{\p\over 2}\right) - 
 (2\pi\e)^{2}iL_{i\l}\left({\k\over \e}+{\p\over 2}\right)W_{\l j}(\p,\k) \\[13pt]
\: & \displaystyle +i\int_{\q} \U(\q)\cdot\left(-{\k\over \e}+
{\p\over 2}-\q\delta_{i2}\right)
\psi_{i}\left(
-{\k \over \e}+{\p \over 2}-\q\right)\psi_{j}^{*}\left(-{\k \over \e}-{\p \over
2}\right) \\[13pt]
\: & \displaystyle - i\int_{\q} \U^{*}(\q)\cdot
\left(-{\k\over \e}-{\p\over 2}-\q\delta_{j2}\right)
\psi_{i}\left(
-{\k \over \e}+{\p \over 2}\right)\psi_{j}^{*}\left(-{\k \over \e}-{\p \over
2}-\q\right) \\[13pt]
\: & \displaystyle + i\sqrt{\e}\int_{\q} \d\U(\q)\cdot
\left(-{\k\over \e}+{\p\over
2}-{\q\over \e}\delta_{i2}\right)\psi_{i}\left(
-{\k \over \e}+{\p \over 2}-{\q\over \e}
\right)\psi_{j}^{*}\left(-{\k \over \e}-{\p \over
2}\right) \\[13pt]
\: & \displaystyle - i\sqrt{\e}
\int_{\q} \d\U^{*}(\q)\cdot\left(-{\k\over \e}
-{\p\over 2}-{\q\over \e}\delta_{j2}\right)
\psi_{i}\left(
-{\k \over \e}+{\p \over 2}\right)\psi_{j}^{*}\left(-{\k \over \e}-{\p \over
2}-{\q\over \e}\right)
\end{array}
\label{WDOT1}
\end{equation}

To rewrite the above expression as a function of $W_{ij}$ only, we
relabel appropriately, {\it e.g.,}
\begin{equation}
\begin{array}{rl}
\displaystyle -{\k \over \e}-{\p \over 2} & \displaystyle  
=-{\k'\over \e}-{\p'\over 2} \\[13pt]
\displaystyle -{\k\over \e}+{\p\over 2}-\q & \displaystyle = 
-{\k'\over \e}+{\p'\over 2}
\end{array}
\end{equation}
for the third term on the right
hand side of (\ref{WDOT1}). Similarly relabelling for all relevant
terms yields the integral equation (\ref{WDOT2}).

The $O(\e^{-1/2})$ terms of (\ref{WDOT2}) determine $\W_{1/2}$.
Decomposing
\begin{equation}
\displaystyle \W_{1/2}(\p,\xib,\k) \equiv
\sum_{\tau,\tau'=\pm}
a^{(1/2)}_{\tau,\tau'}(\p,\xib,\k)
\b_{\tau}(\k_{-})\b_{\tau'}^{\dagger}(\k_{+})
\end{equation}
and substituting into (\ref{SQRTE}) we find the coefficients
$a_{\tau,\tau'}^{(1/2)}$, where in this case
$a_{+-}^{(1/2)},a_{-+}^{(1/2)} \neq 0$.  Due to the nonlocal nature of
the third term on the right of (\ref{SQRTE}), we must first inverse
Fourier transform the slow wavevector variable back to $\x$.

To extract the $O(\e^{0})$ terms from (\ref{WDOT2}) we need to expand
${\bf L}$ to order $\e^{0}$, the ${\bf L}_{1}$ term.  Similarly, the
terms $\W(\p-\q,\xib,\k\pm\e\q/2)$ must be expanded:
\begin{equation}
\W(\p-\q,\xib,\k\pm\e\q/2) = 
\W(\p-\q,\xib,\k) \pm {\e\over 2}\q\cdot\nabla_{\k}
\W(\p-\q,\xib,\k) + O(\e^{2}).
\end{equation}
The $\e\q\cdot\nabla_{\k} \W(\p-\q,\xib,\k)$ terms combine with the
$-\e^{-1}\U(\q)\cdot\k_{-}+\e^{-1}\U(\q)\cdot \k_{+}$ terms from the
third and fourth terms in (\ref{WDOT2}) to give the third term on the
right of Eq.  (\ref{ORDER0}). The $\d\U$-dependent, order $\e^{0}$
terms (the sixth, seventh, and eighth terms on the right side of
(\ref{ORDER0})) come from collecting

\begin{equation}
\pm\sqrt{\e}\d\U(\q)\cdot\left(-{\k\over
\e}\pm{\xib\over 2\e}\right)\sqrt{\e}\W_{1/2}
(\p,\xib-\q,\k\pm\q/2)
\end{equation}
from the last two terms in (\ref{WDOT2}). The ensemble averaged time
evolution of the Wigner amplitude $a_{\sigma}(\x,\k)$ can be
succinctly written in the form:
\begin{equation}
\begin{array}{l}
\dot{a}_{+}(\x,\k,t) -\nabla_{\x}\omega(\x,\k)\cdot\nabla_{\k}a_{+}(\x,\k,t)
+ \nabla_{\k}\omega(\x,\k)\cdot\nabla_{\x}a_{+}(\x,\k,t)\\[13pt]
\:\hspace{2in} = \displaystyle \Sigma_{+,\mu}(\k)a_{\mu}(\x,\k,t) +
\int_{\q}\sigma_{+,\mu}(\q,\k)a_{\mu}(\x,\k,t).
\end{array}
\end{equation}

Using the form for $\W_{0}$ found from (\ref{ZERO}) in (\ref{SQRTE})
to find $\W_{1/2}$, we substitute into (\ref{ORDER0}) to find
(\ref{RESULT1}), the transport equation for one of the diagonal
intensities of the Wigner distribution. We have explicitly used
eigenbasis orthonormality $\b_{\tau}^{\dagger}(\k)\cdot\c_{\tau'}(\k)
=\d_{\tau,\tau'}$ and the fact that $a_{-}(\x,\k,t) =
a_{+}(\x,-\k,t)$.

\section{Derivation of the diffusion equation}

The derivation of diffusion of water wave action is outlined below and
follows the established mathematical treatment of \cite{LK} and
\cite{dlen6}. For simplicity we assume that the flow $\U$ is constant
and small enough so that for a considered range of frequencies, the
relation $\omega(\q)+\omega(\k)=0$ is never satisfied for any $\k$ and
$\q \neq {\bf 0}$.  The diffusion approximation is valid after long
times and large distances of propagation ${\bf X}$(see Fig.
\ref{FIG1}) such that the wave has multiply scattered and its dynamics
are determined by a random walk.  We therefore rescale time and space
as
\begin{equation}
  \label{eq:rescal}
  \tilde t = \frac{t}{\eps^2}, \qquad \tilde \x=\frac{\x}{\eps}.
\end{equation}
The small parameter $\eps$ in this further rescaling represents the
transport mean free path $\ell_{\mbox{{\it mfp}}}$ and not the
wavelength as in the initial rescaling used to derive the transport
equation.  We drop the tilde symbol for convenience and rewrite the
transport equation in the new variables:
\begin{equation}
  \label{eq:transresc}
  \dot{a}_\eps(\x,\k,t) 
+ \frac{1}{\eps}\nabla_\k \omega (\k)\cdot\nabla_\x a_\eps(\x.\k,t)
=\frac{1}{\eps^2} \int_{\q} {\cal
Q}(\q,\k)(a_\eps(\x,\q,t)-a_\eps(\x,\k,t))
\delta(\omega(\q)-\omega(\k)),
\end{equation}
with obvious notation for ${\cal Q}(\q,\k)$. Since the frequency is
fixed, the equation is posed for $\k$ satisfying
$\omega(\k)=\omega_{0}$.  The transport equation assumes the form
(\ref{eq:transresc}) because the scattering operator is conservative.
Since $\U\ne0$, wave action is transported by the flow, and diffusion
takes place on top of advection.  Therefore, we introduce the main
drift $\bar\U$, which will be computed explicitly later, and define
the drift-free unknown $\tilde a_\eps(\x,\k,t)$ as
\begin{equation}
  \label{eq:tildea}
  \tilde a_\eps(\x,\k,t)=a_\eps(\x+\frac{\bar\U}{\eps}t,\k,t).
\end{equation}
It is easy to check that $\tilde a_\eps$ satisfies the same transport
equation as $a_\eps$ where the drift term $\nabla_\k \omega$ has been
replaced by $\nabla_\k \omega - \bar \U$.

We now derive the limit of $\tilde a_\eps$ as $\eps\to0$. Consider the
classical asymptotic expansion
\begin{equation}
  \label{eq:expdiff}
  \tilde a_\eps=\tilde a_0+\eps \tilde a_1 + \eps^2 \tilde a_2 + \ldots.
\end{equation}
Upon substitution into (\ref{eq:transresc}) and equating like powers
of $\eps$, we obtain at order $\eps^{-2}$, for fixed frequency
$\omega_0$,
\begin{equation}
  \label{eq:order-2}
  \int_{\q} {\cal Q}(\q,\k)(\tilde a_0(\x,\q,t)-\tilde a_0(\x,\k,t))
\delta(\omega(\q)-\omega_0)=0.
\end{equation}
It follows from the Krein-Rutman theory (\cite{dlen6}) that $\tilde
a_0$ is independent of $\q$. At order $\eps^{-1}$, we obtain
\begin{equation}
  \label{eq:order-1}
  (\nabla_\k \omega(\k)-\bar\U) \cdot\nabla_\x \tilde a_0 
 =\int_{\q} {\cal Q}(\q,\k)(\tilde a_1(\x,\q,t)-\tilde a_1(\x,\k,t))
        \delta(\omega(\q)-\omega_0).
\end{equation}
The compatibility condition for this equation to admit a solution
requires both sides to vanish upon integration over
$\delta(\omega(\k)-\omega_{0})d\k$.  Therefore, $\bar\U$ satisfies
\begin{equation}
  \label{eq:barU}
  \bar\U=\frac{1}{V_{\omega_0}}\int_\k \nabla_\k \omega(\k)
         \delta(\omega(\k)-\omega_{0}), \qquad \mbox{ where } \qquad
  V_{\omega_0}= \int_\k \delta(\omega(\k)-\omega_{0}).
\end{equation}
Once the constraint is satisfied, we deduce from Krein-Rutman theory
the existence of a vector-valued mean zero corrector $\bchi$ solving
\begin{equation}
  \label{eq:chi}
   (\nabla_\k \omega(\k)-\bar\U)
 =\int_{\q} {\cal Q}(\q,\k)(\bchi(\q)-\bchi(\k))
        \delta(\omega(\q)-\omega_{0}) \equiv {\cal L}\bchi.
\end{equation}
There is no general analytic expression for $\bchi$, which must in
practice be solved numerically. This is typical of problems where the
domain of integration in $\q$ does not have enough symmetries (cf.
\cite{allbal,balsiam}). We now have
\begin{equation}
  \label{eq:a_1}
  \tilde a_1(\x,\k,t) = \bchi(\k)\cdot\nabla_\x \tilde a_0(\x,t).
\end{equation}
It remains to consider $O(\eps^{0})$ in the asymptotic expansion. This
yields
\begin{equation}
  \label{eq:order0}
  \dot{\tilde a}_0 + (\nabla_\k \omega(\k)-\bar\U)\cdot\nabla_\x \tilde a_1
= {\cal L} (\tilde a_2).
\end{equation}
The compatibility condition, obtained by integrating both sides over
$\k$, yields the wave action diffusion equation
\begin{equation}
  \label{eq:diffusion}
  \dot{\tilde a}_0 - \nabla_\x \cdot{\bf D}\cdot\nabla_\x \tilde a_0 = 0,
\end{equation}
where the diffusion tensor is given by
\begin{equation}
  \label{eq:coeffD}
  {\bf D} = -\frac{1}{V_{\omega_0}} \int_\k (\nabla_\k \omega(\k)-\bar\U)
   \bchi^T(\k)=-
    \frac{1}{V_{\omega_0}} \int_\k {\cal L}(\bchi) \bchi^T(\k).
\end{equation}
The second form shows that ${\bf D}$ is positive definite since ${\cal
  L}$ is a nonpositive operator. The formal asymptotic expansion can
be justified rigorously using the techniques in \cite{dlen6}. As
$\eps\to0$, we obtain that the error between $\tilde a_\eps$ and
$\tilde a_0$ is at most of order $\eps$. Therefore, we have that
$a_\eps$ converges to $a_0$ satisfying the following drift-diffusion
equation
\begin{equation}
  \label{eq:driftdiff}
    \dot{a}_0 +\frac{\bar\U}{\eps} \cdot\nabla_\x a_0 
 - \nabla_\x \cdot{\bf D}\cdot \nabla_\x a_0 = 0,
\end{equation}
with suitable initial conditions. Equation (\ref{eq:driftdiff}) is the
coordinate-scaled version of (\ref{eq:diffa0}).

\end{document}